\documentclass[twocolumn,showpacs,preprintnumbers,amsmath,amssymb,superscriptaddress]{revtex4-1}

\usepackage{upgreek}
\usepackage{graphicx}
\usepackage{dcolumn}
\usepackage{bm}

\begin{document}

\title{Two-Color Photon Correlations of the Light Scattered by a Quantum Dot}

\author{M. Peiris}
\affiliation{Physics Department, University of South Florida, Tampa, Florida}
\author{B. Petrak}
\affiliation{Physics Department, University of South Florida, Tampa, Florida}
\author{K. Konthasinghe}
\affiliation{Physics Department, University of South Florida, Tampa, Florida}
\author{Y. Yu}
\affiliation{State Key Laboratory of Superlattices and Microstructures, Institute of Semiconductors, Chinese Academy of Sciences, Beijing, China}
\affiliation{Synergetic Innovation Center of Quantum Information and Quantum Physics, University of Science and Technology of China, Hefei, China}

\author{Z. C. Niu}
\affiliation{State Key Laboratory of Superlattices and Microstructures, Institute of Semiconductors, Chinese Academy of Sciences, Beijing, China}
\affiliation{Synergetic Innovation Center of Quantum Information and Quantum Physics, University of Science and Technology of China, Hefei, China}

\author{A. Muller}
\email{mullera@usf.edu}

\affiliation{Physics Department, University of South Florida, Tampa, Florida}

\date{\today}

\begin{abstract}
Two-color second-order correlations of the light scattered near-resonantly by a quantum dot were measured by means of spectrally-filtered coincidence detection. The effects of filter frequency and bandwidth were studied under monochromatic laser excitation, and a complete two-photon spectrum was reconstructed. In contrast to the ordinary one-photon spectrum, the two-photon spectrum is asymmetric with laser detuning and exhibits a rich structure associated with both real and virtual two-photon transitions down the ``dressed states'' ladder. Photon pairs generated via virtual transitions are found to violate the Cauchy-Schwartz inequality by a factor of 60. Our experiments are well described by the theoretical expressions obtained by del Valle {\it et al.} via time-and normally-ordered correlation functions.
\end{abstract}

\pacs{78.67.Hc, 78.47.--p, 78.55.Cr}
                          
\maketitle

Quantum particles generated in pairs are an essential resource in quantum information science and a unique testbed for the investigation of quantum mechanical paradoxes \cite{einstein}. There are currently two major approaches for generating pairs of photons which are correlated strongly enough to violate classical inequalities such as the Cauchy-Schwartz or Bell's inequalities \cite{reid1986,bell}. One relies on nonlinear parametric processes like spontaneous down-conversion \cite{kwiat1995nhi} or four-wave mixing \cite{kolchin2006}. The other uses multilevel atomic cascades, as in the pioneering experiments of Aspect {\it et al.} \cite{aspect1982etb}, or more recently, in biexcitonic decays of quantum dots (QDs) \cite{akopian2006,young2006ift,hafenbrak2007}. 

In light of the complexities associated with obtaining paired photons in a {\it multilevel} system, it is natural to ask under which conditions a {\it two-level} system may generate photon pairs correlated strongly enough to violate classical inequalities \cite{munoz2014vci}. It is well-known that spectrally filtering the light scattered by strongly driven two-level atoms yields correlated pair emission \cite{aspect1980,dalibard1983csr,nienhuis1983,knoll1984,arnoldus1984,knoll1986qns,knoll1986,schrama1992icb,nienhuis1993scr,artoni1999pnl,joosten2000isf}. In QDs this cascaded emission has been investigated recently by using a Michelson interferometer to separate different components of the spectral Mollow triplet \cite{ulhaq2012csp}. In this way the sequential emission of photons in pairs emanating from the two Mollow triplet sidebands has been demonstrated.

In general, scattering of photon pairs from a strongly driven two-level system may occur via numerous pathways, and any photon with one color and emission time may be correlated, to some degree, to another photon with possibly different color and emission time. For the purpose of describing such correlations, del Valle {\it et al.} have introduced a ``two-photon spectrum'' (TPS) as an extension of the ordinary one-photon spectrum \cite{delvalle2012tff}. While the latter simply measures the probability of detecting a photon of frequency $\omega$ (obtained experimentally by recording the transmission of light through a frequency-tunable filter), the TPS measures the probability of detecting one photon of frequency $\omega_1$ at time $T_1$ and another of frequency $\omega_2$ at time $T_2$. Experimental measurement of the TPS requires frequency-resolved coincidence detection, such as is obtained by placing tunable filters in front of each detector of a Hanbury-Brown and Twiss (HBT) setup and histogramming photon arrival times \cite{cresser1987icf}. The theoretical framework for the TPS calculation rigorously yields the frequency and bandwidth dependent two-color correlations which are required to determine whether or not classical inequalities are violated \cite{munoz2014vci}.

We report here the measurement of the TPS of the light near-resonantly scattered by a QD exposed to a strong monochromatic laser. The 2D TPS maps reveal intricate and unexpected features of two-photon cascade emission such as transitions proceeding via virtual intermediate states showing particularly strong correlations, significantly violating the Cauchy-Schwartz inequality. Furthermore, we evidence the asymmetric nature of the TPS under laser detuning as well as the effect of filter bandwidth on the correlations at a level at which Rabi oscillations can be resolved. The TPS measurement provides new opportunities for the characterization of quantum optic pathways which could help improve our understanding of a variety of systems \cite{gonzalez2013tps}.

\begin{figure*}[t!]
\includegraphics[width=6.8in]{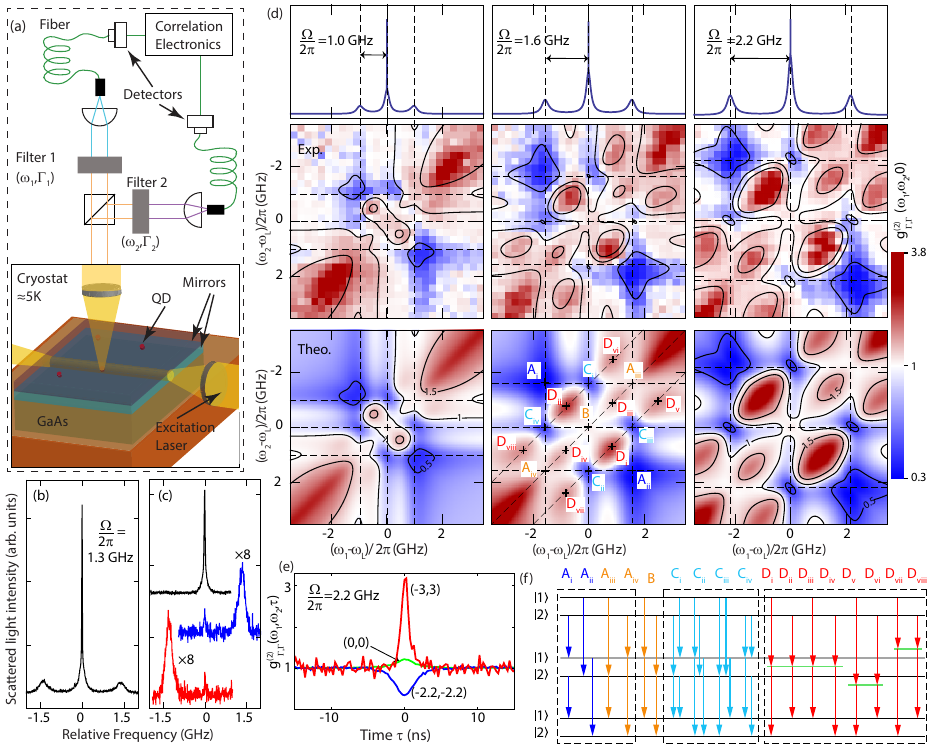}
\caption{\label{tpspec}  (a) Experimental setup. (b) Unfiltered Mollow triplet for $\Omega /2 \pi$=1.3 GHz. (c) From top to bottom: isolated central peak, blue and red sidebands, respectively ($\Gamma /2 \pi$=0.5 GHz). (d) Experimental and theoretical TPS for the Rabi frequencies indicated, with filter bandwidths of $\Gamma_1 /2 \pi=\Gamma_2 /2 \pi$=0.5 GHz. (e) Experimental correlations for $\Omega /2 \pi$=2.2 GHz labeled by $( \frac{\omega_1-\omega_L}{2\pi}, \frac{\omega_2-\omega_L}{2\pi}$) in units of GHz. (f) Dressed-state diagram illustrating the transitions labelled in the middle theory panel of part (d).}
\end{figure*}

We used InAs QDs grown by molecular beam epitaxy (see Ref. \cite{konthasinghe2012cvi} for sample details) in a cryogenic orthogonal excitation/detection setup, at a cryostat base temperature of 5 K [Fig. 1(a)]. With this geometry the light scattered by a single QD exposed to a tunable continuous-wave laser is collected efficiently, free of unwanted background laser scattering \cite{muller2007rfc}. We are interested in the situation in which two filters, tunable both in their resonance frequencies, $\omega_1$ and $\omega_2$, as well as in their bandwidths, $\Gamma_1$ and $\Gamma_2$, are placed before the detectors of a HBT setup \cite{michler2000} receiving the scattered light. Figure 1(b) displays the scattered light (one-photon) spectrum for a single QD obtained using a high resolution ($\approx$ 20 MHz) scanning Fabry-Perot interferometer at a Rabi frequency of $\Omega/2\pi$=1.3 GHz. The functionality of the filters is illustrated in Fig. 1(c). Each filter could be tuned continuously; in particular it could select the Mollow triplet red or blue sidebands, its central peak, as shown, or any other frequency window. The filters' long-term stability has been verified in separate measurements.

In order to record the TPS associated with the QD scattered light, photon arrival times were histogrammed for a matrix of filter frequencies $(\omega_1,\omega_2)$, using a fixed filter bandwidth $\Gamma_1=\Gamma_2=\Gamma$. The QD resonance frequency will be denoted by $\omega_0$ while the laser frequency will be denoted by $\omega_L$. We define the laser detuning as $\delta=\omega_L-\omega_0$.  Figure 1(d) shows the results of the TPS measurement under coincidence detection, i.e., for a correlation time $\tau=T_2-T_1=0$, and with $\delta=0$ and $\Gamma /2 \pi$=0.5 GHz. Only the Rabi frequency was changed between each panel (increasing from left to right). The theoretical one-photon spectrum is plotted above the TPS for ease of comparison and identification of spectral features. Figure 1(e) shows several sample raw correlation functions from which the images were constructed. Further details are provided in the form of supplemental material.

For a qualitative understanding we treat the resonantly-driven QD as an ideal two-level system with ground state $|g\rangle$, excited state $|e\rangle$, and radiative decay rate $\kappa$. By diagonalizing the Hamiltonian of the coupled system (QD+laser field), the ``dressed states'', $|1\rangle= c|g\rangle -s|e\rangle$ and  $|2\rangle=s|g\rangle + c|e\rangle $,  are obtained for each rung of the dressed states ``ladder'' \cite{aspect1980,nienhuis1993scr}. The amplitudes of the eigenstates are given by $ c=\sqrt{(\Omega'+\delta)/2\Omega'}$ and $s=\sqrt{(\Omega'-\delta)/2\Omega'}$, with $\Omega'^{2}=\Omega^2+\delta^2$ \cite{nienhuis1993scr}. The dressed states picture is ideally suited for visualizing the two-photon cascade [Fig. 1(f)]. In this picture, the lowest coincidence rate, i.e., highest degree of photon anti bunching, is predicted to occur near $(\omega_1-\omega_L,\omega_2-\omega_L)=(\pm \Omega, \pm \Omega)$ [labelled A$_i$ and A$_{ii}$], due to disconnected decay paths. These correspond to filtering of like sidebands. On the other hand, a high coincidence rate is expected to occur near $(\omega_1-\omega_L,\omega_2-\omega_L)=(\pm \Omega, \mp \Omega)$ [A$_{iii}$ and A$_{iv}$]. These are associated with the cascaded alternating sideband emission \cite{nienhuis1993scr}. Other notable pathways are those associated with filtering the central line, which interfere to yield uncorrelated statistics (B), and those associated with filtering the central line and one of the sidebands, which interfere to yield partial anti bunching (C$_{i}$-C$_{iv}$) \cite{dalibard1983csr}.

\begin{figure}[b]
\includegraphics[width=3.4in]{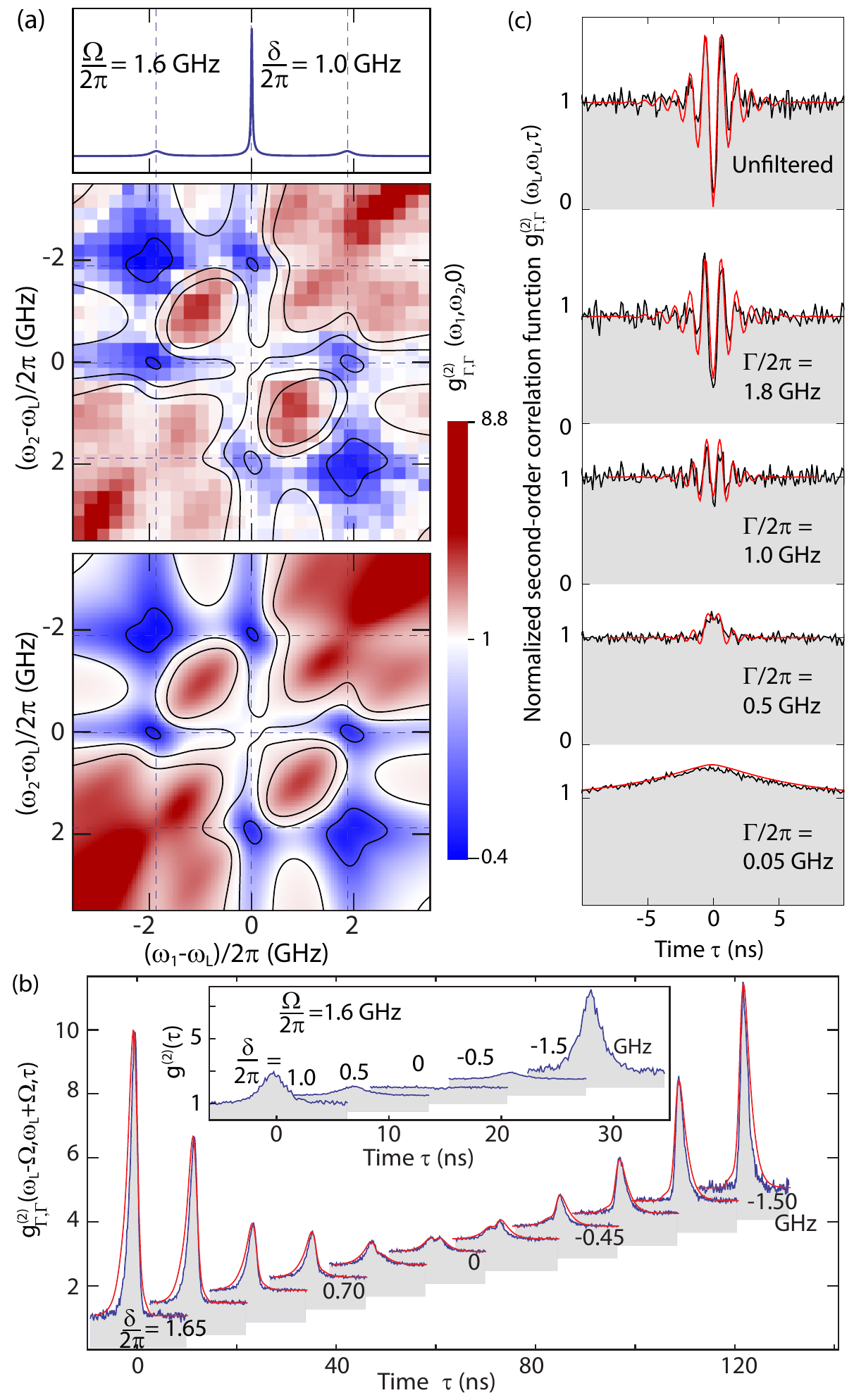}
\caption{\label{tpspectdet} (a) Experimental and theoretical TPS under detuning $\delta/2 \pi$ = 1.0 GHz, with $\Omega /2 \pi$ = 1.6 GHz. (b) Experimental (black) and theoretical (red) photon correlations between red and blue sidebands ($\Omega /2 \pi$ = 1.6 GHz) for different laser detuning ($\delta /2 \pi$= 1.65, 1.40, 0.85, 0.70, 0.30, 0, -0.15, -0.45, -0.75, -1.20 and -1.50 GHz), offset for clarity with the shaded region indicating the zero. Inset: corresponding photon correlation measurements on both Mollow triplet sidebands for a range of detunings as indicated. (c) Experimental (black) and theoretical (red) photon correlations on the central Mollow triplet peak for a range of filter bandwidths ($\Gamma_1  =\Gamma_2=\Gamma $) as indicated.}
\end{figure}

The features described above are well documented \cite{schrama1992icb,nienhuis1993scr,ulhaq2012csp}, and are clearly seen in the TPS of Fig. 1(d) (at the 9 intersections of the dashed gridlines). However, they do not form a dominant pattern. Rather, the regions with the highest coincidence rates are generally located along anti-diagonal lines defined by $\omega_1+\omega_2-2\omega_L=\pm\Omega$ and $\omega_1+\omega_2-2\omega_L=0$, with strong maxima (photon bunching) observed on these lines at points near which $\omega_{1,2}-\omega_L\approx -3\Omega/2,-\Omega/2,\Omega/2,3\Omega/2$. For these two-photon decays, the cascade proceeds via a virtual state, i.e., through an intermediate state which is not an eigenstate of the system. Figure 1(f) depicts several of these transitions (D$_{i}$-D$_{viii}$).
 
For the purpose of quantitatively describing the TPS measurements of Fig. 1(d), del Valle {\it et al.} have introduced the quantity $S^{(2)}_{\Gamma_1 \Gamma_2}(\omega_1,T_1,\omega_2,T_2)=\frac{\Gamma_1 \Gamma_2}{(2\pi)^2}\int\int_{-\infty}^{T_1}dt'_1dt'_4e^{-\frac{\Gamma_1}{2}(T_1-t'_1)}e^{-\frac{\Gamma_1}{2}(T_1-t'_4)} e^{i\omega_1(t'_4-t'_1)}\times$ $\int\int_{-\infty}^{T_2}dt'_2dt'_3 e^{-\frac{\Gamma_2}{2}(T_2-t'_2)}e^{-\frac{\Gamma_2}{2}(T_2-t'_3)} e^{i\omega_2(t'_3-t'_2)}\times$ $\langle\mathcal{T}_{-}[{a^{\dagger}(t'_1)}{a^{\dagger}(t'_2)}]\mathcal{T}_{+}[a(t'_3)a(t'_4)]\rangle$, where $\mathcal{T}_{-}$ and $\mathcal{T}_{+}$ are time-ordering operators \cite{delvalle2012tff}, and $a^\dag$ and $a$ are the photon creation and annihilation operators, respectively. After normalizing by the one-photon time-dependent power spectra $S^{(1)}_{\Gamma_1}(\omega_1,T_1)$ and  $S^{(1)}_{\Gamma_2}(\omega_2,T_2)$, the time and frequency resolved physical TPS is given by
\begin{equation}
g^{(2)}_{\Gamma_1,\Gamma_2}(\omega_1, \omega_2, \tau) = \frac{S^{(2)}_{\Gamma_1,\Gamma_2}(\omega_1, T_1,\omega_2, T_2)}{S^{(1)}_{\Gamma_1}(\omega_1,T_1) S^{(1)}_{\Gamma_2}(\omega_2,T_2)}\Bigg|_{T_2-T_1=\tau} \mathrm{.}
\label{eq:tps}
\end{equation}
In the supplemental material of Ref. \cite{delvalle2012tff} it is shown how this quantity can be calculated in a general context. Here we have evaluated it numerically for the case of a radiatively-broadened two-level system (radiative decay rate of $\kappa/2\pi$=0.2 GHz). It is represented below the experimental TPS in Fig. 1(d). A convolution with the detectors' instrument response ($\Gamma_\mathrm{det}^{-1}$=350 ps) function has also been applied for accurate comparison with experiments.

As can be seen, all major features are predicted well by Eq. (\ref{eq:tps}). Theoretical contours have been overlaid onto the experimental measurements for highlighting purposes. Crucially, the virtual transitions D$_{i}$-D$_{viii}$ of Fig. 1(f) can be reproduced. These have been first identified through the inspection of the Jaynes-Cummings model in Ref. \cite{gonzalez2013tps}, where they were termed ``leapfrog transitions''.

The detuning provides another interesting control parameter for the investigation of the TPS. In Fig. 2(a) the TPS has been measured in the presence of a laser detuning $\delta/2\pi$=1.0 GHz. Despite the moderate magnitude of this detuning, a mirror asymmetry in the TPS is readily seen relative to the central anti-diagonal, in contrast to the strict symmetry of the one-photon spectrum under detuning. Again, the theoretical calculation based on Eq. (\ref{eq:tps}) [bottom map of Fig. 2(a)] provides close agreement with experiment.

The laser detuning also controls the superposition amplitudes $c$ and $s$ of the dressed states. Under large positive detuning ($|c|\gg|s|$) the branching ratios are such that transitioning into state $|1\rangle$ is favored over transitioning into state $|2\rangle$. Thus in steady-state the system is found predominantly in state $|1\rangle$ and the emission of a photon at the blue sideband is likely to be followed by the emission of a photon at the red sideband [A$_{iii}$ in Fig. 1(f)]. This asymmetric time sequence is clearly visible in the data of Fig. 2(b). Likewise, when the detuning is negative then the red sideband photon emission is likely to be followed by a blue sideband photon emission [A$_{iv}$ in Fig. 1(f)]. On resonance A$_{iii}$ and A$_{iv}$ are equally likely to occur and thus the resulting correlation function is symmetric, with a dip at $\tau=0$ due to their interference \cite{dalibard1983csr}. Note that the effect of spectral diffusion \cite{santori2002} has been accounted for in the theory via an averaging over a 1 GHz wide distribution of random detunings as described in Ref. \cite{konthasinghe2012cvi}. We have also verified that our results agree with those reported in Ref. \cite{ulhaq2012csp}, for the correlations observed when the Mollow triplet is stripped of its central peak. This case is shown in the inset of Fig. 2(b), where one sideband was selected by each filter and the two signals were subsequently recombined at a beam splitter before histogramming.

We also explored the effect of reduced filter bandwidth on the two-photon correlations. Here we fixed the frequency of each filter to select the central peak from the Mollow triplet. The correlations were then recorded for increasingly smaller filter bandwidths, as shown in Fig. 2(c). Here the detector time resolution ($\Gamma_\mathrm{det}^{-1}\approx$ 80 ps) was such that Rabi oscillations are seen when no filter is present [uppermost trace in Fig. 2(c)]. A deconvolution with the detectors' instrument response function has been applied to the data. The red continuous traces in Fig. 2(c) correspond to the theoretical correlations [Eq. (\ref{eq:tps})]. These follow closely the experimental data as the filter bandwidth is reduced. Note that the effect of the reduced filter bandwidth is not equivalent to a slower detector response (which would result simply in a removal of high-frequency components of the Rabi oscillations and a ``flattening'' of the correlations). Here the reduced filter bandwidth eventually leads to increased coincidence rates, an effect well-known from the quantum treatment of filter transfer functions  \cite{artoni1999pnl,joosten2000isf}. In particular, regardless of the specific photon generation mechanism, a filter with a bandwidth smaller than the light's bandwidth unavoidably introduces quantum noise \cite{knoll1986qns}, i.e., it will cause photon bunching to occur. This bunching can be understood as being the result of a constructive multiphoton interference process originating in photon indistinguishability \cite{santori2004,ates2012}. Had the TPS of Figs. 1,2 been recorded with a filter bandwidth much less than the emitter bandwidth $\kappa$, then this {\it indistinguishability bunching} would have appeared as a sharp center-diagonal line \cite{gonzalez2013tps}.

\begin{figure}[t]
\includegraphics[width=3.4in]{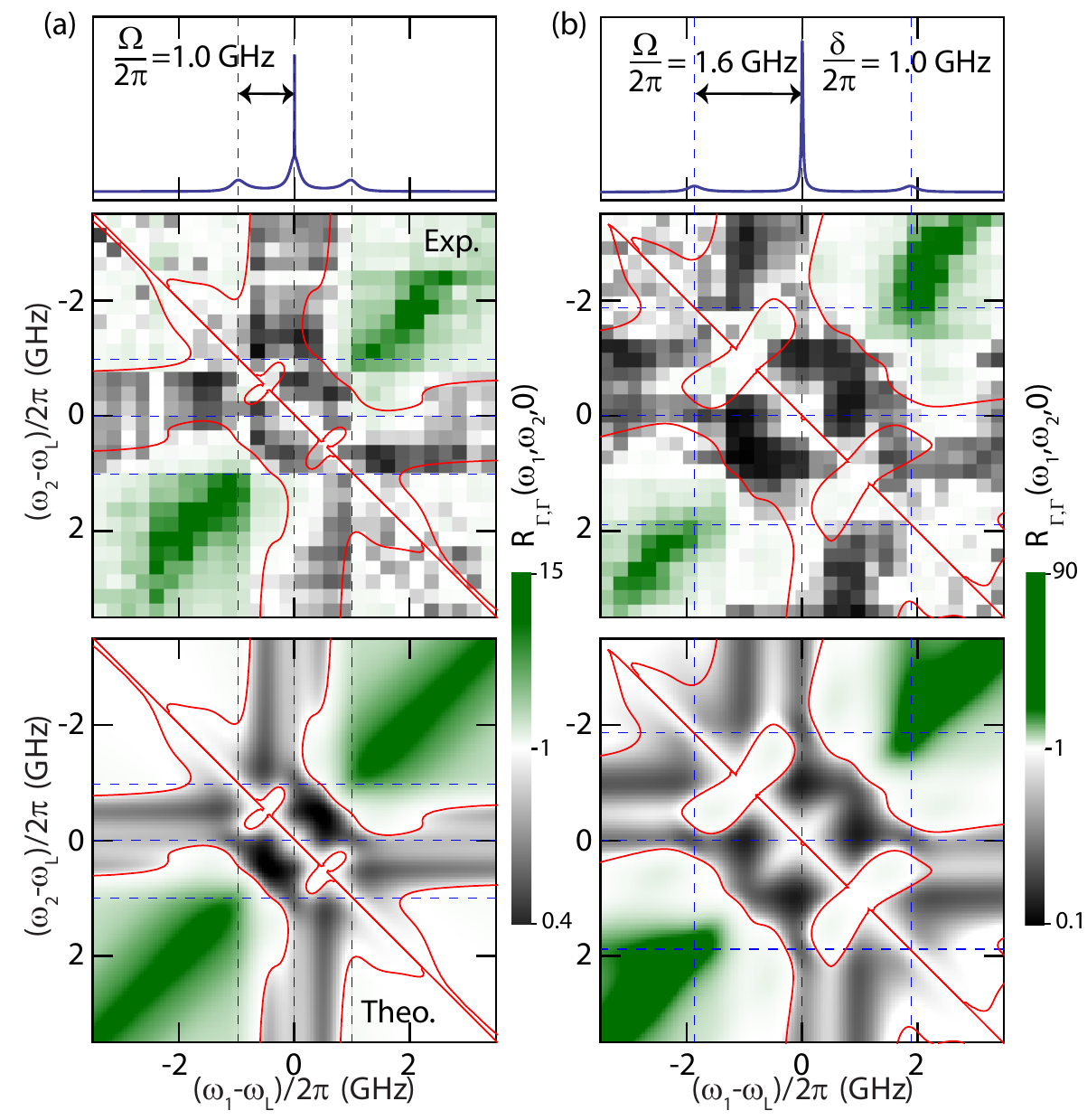}
\caption{\label{detuning} (a) Map of the Cauchy-Schwartz criteria $R$ for with no laser detuning. (b) Same as in (a) but with a laser detuning of $\delta/2\pi$=1.0 GHz.}
\end{figure}

Lastly we examine the conditions under which the scattered photon pairs violate the Cauchy-Schwartz inequality. In quantum optics, for two electromagnetic modes, this inequality is written in the form of the ratio $R$ of the square of cross-correlations over the product of auto-correlations between the modes at $\tau=0$, i.e, using the notation of Eq. (\ref{eq:tps}) \cite{munoz2014vci},
\begin{equation}
\begin{split}
R_{\Gamma_1,\Gamma_2}(\omega_1, \omega_2,0)= \\
\left[g^{(2)}_{\Gamma_1,\Gamma_2}(\omega_1, \omega_2, 0)\right]^2&/\left[g^{(2)}_{\Gamma_1,\Gamma_1}(\omega_1, \omega_1, 0) g^{(2)}_{\Gamma_2,\Gamma_2}(\omega_2, \omega_2, 0)\right] \mathrm{.}
\end{split}
\label{eq:csi}
\end{equation}
In Fig. 3 the experimentally obtained $R$ is plotted as a function of the filter frequencies. The theoretical value is represented in the bottom panels. On the chosen color scale, green indicates a violation ($R>1$). As can be seen, significant violations occur only for pairs emitted in the Mollow triplet sideband tails, unless a laser detuning is present [Fig. 3(b)], in which case a violation is also observed for filters set exactly at opposite sidebands. The largest experimental value seen is $R\approx$60.

In conclusion, we have measured the two-photon spectrum of the light resonantly scattered by a QD. Its distinctive features, such as the visualization of virtual transitions, asymmetry under laser detuning, and filter bandwidth-dependent effects have been identified. Although a refined theoretical description may be obtained by including, e.g., phonon-scattering---known to play an important role in QD exciton dephasing \cite{forstner2003pad,Ahn2005,ramsey2010der}---our observations agree well with the theoretical model of del Valle {\it et al.} \cite{delvalle2012tff}. The TPS measurement may provide new insights into more complex systems, such as the two-level atom dressed by a cavity vacuum field \cite{kasprzak2010uad}, as well as coupled quantum systems. Finally, our results also suggest that a demonstration of Bell's inequalities, as predicted theoretically \cite{munoz2014vci}, should be possible experimentally.

The authors acknowledge financial support from the National Science Foundation (NSF grant No. 1254324) and the National Natural Science Foundation of China (Grant No. 90921015).

\end{document}